\documentclass{elsart}
\usepackage{psfig}
\usepackage{natbib}

\begin{document}

\runauthor{Collin and Joly}


\begin{frontmatter}

\title{The Fe~II problem in NLS1s}

\author[PSU]{Suzy Collin, Monique Joly}
\address[PSU]{DAEC Observatoire de Paris-Meudon 92195 Meudon 
cedex, France}

\begin{abstract}
	For more than twenty years, strong FeII emission lines have 
been observed
in Active Galactic Nuclei and in particular in Narrow Line Seyfert~1
galaxies. A quick overview of the observations and of the models 
proposed to interpret the
Fe~II spectrum is given. The influence of atomic data and of physical 
parameters are discussed, and
it is shown that the strengths of the Fe~II lines cannot be explained 
in the framework of photoionization models. A non-radiative 
heating, for instance due to shocks, with an overabundance of iron, 
can help to solve the problem. A comparison
with other objects emitting intense Fe~II lines favors also the 
presence of strong outflows and shocks. We suggest some issues in 
the context of AGN evolution. \end{abstract}

\begin{keyword}
galaxies: active; quasars: emission lines; Fe~II \end{keyword}

\end{frontmatter}


\section{Introduction}

Emission lines of Fe~II have been
identified in the spectra of the first quasars by Greenstein and 
Schmidt (1964),
and soon after by Wampler and Oke (1967). In 1968, Sargent 
identified several
strong Fe~II blends in the active galaxy 1 Zw~1. The importance of 
Fe~II in astrophysical objects prompted a meeting gathering 
specialists from different fields, organized in 1979 in Madrid by 
Mike Penston, and another one organized in 1987 in Capri by Viotti, 
Vittone, and Friedjung. In the meantime, several
theoretical works aimed at explaining the formation of Fe~II lines 
were
published. Rapidly it was realized that Fe~II lines are more intense 
when the
broad lines are narrow, but this was attributed to the difficulty of 
deblending the lines. Now it is well known that they
are very intense in NLS1s. In 1992 Boroson and Green identified the
optical Fe~II lines as the eigenvector 1 in their statistical study. 

In the next section we recall the main characteristics of the Fe~II 
atom, and in Section 3, the observational data. In Section 4 we 
discuss the 
processes of line
formation, and we address in Section 5 the question of whether Fe~II in
NLS1s can be explained in the framework of photoionization models. 
In Section 6 we
recall the characteristics of other objects displaying intense Fe~II 
lines, and we suggest a clue to the Fe~II problem which matches well 
the picture of NLS1s.

\section{The Fe~II atom and its spectrum} 

\begin{figure}
\centerline{\psfig{figure=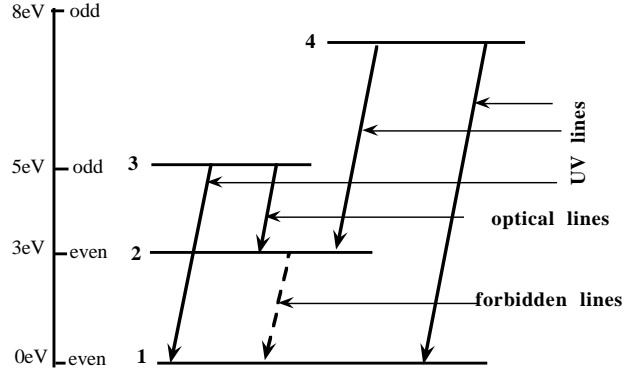,width=0.6\textwidth}} 
\caption{The Fe~II atom}
\label{fig-FeII-atom}
\end{figure}

There are at least three reasons why the Fe~II spectrum is so peculiar. 
First, as the 3d-shell is half-filled, thousands of transitions can 
occur.
Second, FeI is easily ionized in a
weakly
ionized medium (Ionization Potential IP=7.9 eV), while Fe~II is not 
(IP=16.2 eV). And third, a
temperature
$\sim$ 6000 K is sufficient to excite Fe~II. 

The Fe~II atom can be shown schemantically  by 4 levels (see Netzer, 1988) 
producing
four types of lines (cf. Fig. \ref{fig-FeII-atom}): \begin{itemize}
\item Transitions between odd 5eV levels and even ground levels 
produce
permitted UV multiplets in the range 2000 - 3000\AA. The most 
intense
multiplets are UV1, UV2, UV3, UV35, UV36, UV62, UV63 and UV64. 
\item Transitions between odd 9eV and even 3eV levels give rise to 
permitted
multiplets such as UV191 around 1786\AA. \item Transitions 
between even 3eV levels and even ground levels are forbidden
optical multiplets, and are weak or not observed in AGN. \item Transitions 
between odd 5eV levels and even 3eV levels correspond to the
permitted optical multiplets 27, 28 and 29, around 4000\AA, to the 
well-known feature at $\sim$ 4570\AA\ including mainly 
multiplets 37 and 38, to multiplets 48 and 49 around 5300\AA, to 
multiplet 40 at 6516\AA, etc... \end{itemize}
A more complete Fe~II atom is given by the Grotrian diagram of the 
reduced
atom configuration limited to 14 levels (Moore \& Merill, 1968). 

\section{Observations of Fe~II in NLS1s and in quasars} 

In the optical, generally only the feature around 4570\AA\ is 
measured, with
large variations between different authors. There are very few 
measurements in the UV. Moreover models are required to
determine the level of the continuum, owing to the numerous lines 
and the
Balmer continuum (cf. Wills et al., 1985). Large discrepancies arise, 
even
on the identifications of the lines (ex: Fe~III instead of Fe~II is now
identified in the spectrum of 1 Zw1 by Laor et al., 1997). Table 1 
presents the average values of a few important features
measured in the composite QSO spectrum of Francis et al. (1991), in 
the
composite spectrum of X-bright and X-faint QSO of Green (1998), 
and in the
well-known prototype NLS1 galaxy 1 Zw1. In 1 Zw1 the sum of all 
the Fe~II lines is larger than in QSOs, and Fe~IIUV is
$\leq$ Fe~IIopt (Fe~IIUV measures the emission in the range 2000- 
3000\AA, and
Fe~IIopt the emission in the range 3500-6000\AA). Moreover, the 
measured Fe~IIopt is
probably an underestimation as suggested by computations of Joly 
(1987)
which show that Fe~IIopt $\sim$ 10 Fe~II4570. 

\begin{table*}
\caption{Fe~II observations in QSO and in 1 Zw1} \bigskip 
\begin{tabular}{ccccc}
\hline
\noalign{\smallskip}
& 688 QSO & 60 X-bright & 54 X-faint & 1 Zw1 \\ & Francis & 
Green 1998 &
Green 1998 &	\\
&et al. 1991&	&	&	\\
\noalign{\smallskip}
\hline
\noalign{\smallskip}
FeII2400/H$\beta$ & 1.2	& 0.6	& 0.8	& 2	\\ 
FeIIUV/H$\beta$ & 3.3	&	&	& 4.4 \\
FeII4570/H$\beta$ & 0.5	& 0.4	& 0.6	& 1.3-1.7 \\ 
FeIIopt/H$\beta$ & 2.6	&	&	& 6.7 \\
FeIItot/H$\beta$ & 6.	&	&	& 11. \\
\noalign{\smallskip}
\hline
\end{tabular}
\bigskip
\end{table*}

	A number of correlations have been observed: \begin{itemize}
\item A large Fe~II4570/H$\beta$ ratio and a strong equivalent 
width W(Fe~II4570) correlate with a small FWHM(H$\beta$) ratio 
(Boroson et al., 1985, Zheng \& O'Brien, 1990, Zheng \& Keel, 1991). 
\item A large Fe~II4570/H$\beta$ ratio correlates with a weak 
[OIII]/H$\beta$ ratio (Boroson \& Green, 1992, and subsequent 
papers). \item A large Fe~II4570/H$\beta$ ratio correlates with a 
blue asymmetry of H$\beta$
(Boroson \& Green, 1992, and subsequent papers). \item AGN with 
a large Fe~II4570/H$\beta$ ratio have steep IR-X indices (Lawrence 
et al., 1997) and a steep soft X-ray continuum (Wilkes et al., 1987, 
Wang et al., 1996). \item A large Fe~II4570/H$\beta$ ratio 
corresponds to a high soft X-ray variability (Moran et al. 1996)
\item Fe~II is strong in low ionization BAL QSOs (Hartig \& Baldwin, 
1986, Wampler, 1988, Junkharinen et al., 1987, Weyman
et al., 1991, Sprayberry \& Foltz, 1992, Boroson \& Meyers, 1992). 
\item Fe~II is weak in steep spectrum radio sources (Boroson \& Green, 1992).
\item A strong 
W(Fe~II4570) correlates with a large R, the radio core/lobe 
brightness
ratio in radio sources (Joly, 1991).
\item A large Fe~II/[OIII] ratio correlates with a large R (Jackson \& 
Browne, 1991).
\end{itemize}

\section{Formation of the Fe~II spectrum} 

Important improvements were
introduced since the first model computations. \begin{itemize}
\item The number
of levels of the Fe~II model atom has been increased: Collin et al. 
(1986, 1988) and Joly (1981, 1987) used a 14-level atom; Krolik \& 
Kallman (1988) used a 16-level atom; Wills et al. (1985) used 1084 
multiplets, corresponding to 3407 lines;
Verner et al. (1999) used 371 levels producing 68635 lines; Sigut 
and Pradhan (2000) are preparing an atom with 827 levels. \item 
Experimental values for all energy levels are now available;
photoionization and recombination processes are obtained from 
Opacity
Project calculations and transition probabilities and collision 
strengths
are estimated now with less than 30$\%$ uncertainty for strong 
lines. \end{itemize}

	Several possible excitation processes
have been
extensively discussed.
\begin{itemize}
\item Continuum fluorescence (i.e. absorption of continuum
photons by Fe~II transitions, Phillips, 1978, 1979), has been shown 
to be inefficient unless the covering factor of the central source by 
the
BLR is $\sim$1 and the turbulent velocity is very large (Collin et al.
1979, 1980)
\item Line fluorescence has also been considered for exciting high 
levels. Self fluorescence was studied by Netzer \& Wills (1983) as 
well as other line fluorescence. Penston (1988) noticed that 
unexpected UV
multiplets in the symbiotic star RR Telescopi are issued from high 
levels which could be excited by L$\alpha$. A strong 
efficiency requires a large turbulent velocity. Although the 
efficiency of the process can be high (Sigut \& Pradhan, 1998; 
Verner et
al., 1999), it is not able to double the whole Fe~II
flux but it can at least explain the excitation of specific lines (such 
as Fe~II UV191).
\item In fact, collisional excitation is the most efficient process in 
AGN. It requires a high density and T $>$ 5000K as shown by Joly 
(1981, 1987, 1991). \end{itemize}

\section{Models}

\subsection{Photoionization models}

Immediately after the discovery of the presence of Fe~II lines in 
AGN spectra, it was clear that they should be formed in a dense 
medium, comparable to the Broad Line Region (BLR), since Fe~II 
forbidden lines are absent. Therefore people tried to account for the 
Fe~II spectrum in the framework of the BLR photoionization
models developed in the eighties. The first models were 
those of Kwan and Krolik (1981) and of Netzer (1980). Netzer \& 
Wills
(1983) and Wills et al. (1985) introduced a much larger number of 
levels
for Fe~II. Photoinization models computed by Collin et al. (1986, 
1988)
have a small number of levels, but they solve the exact line transfer 
and for some of
their models, the physical conditions in the BLR mimic an accretion 
disk. Krolik \& Kallman (1988) computed photoionization models 
with 
several ionizing continua. Ferland \& Persson (1989) computed CaII 
and Fe~II models with very large thickness and low
density. Sigut \& Pradhan (1998) coupled the computation with 
CLOUDY of the structure of the emission region with a Fe~II atom 
including 3400 transitions, with an exact line transfer of
L$\alpha$ aimed at determining accurately the influence of the 
fluorescence. Finally, Verner et al. (1999) used CLOUDY for an
Fe~II atom of 371 levels.
Note that all these
computations except those of Collin et al. (1986, 1988), and partly 
those of Sigut \&
Pradhan (1998), solve the line transfer with the so-called ``escape 
probability approximation", which is not well adapted to thick 
inhomogeneous media, in
particular to visible permitted Fe~II lines whose transfer is strongly 
linked with that of the (optically thick) Balmer continuum. 
Moreover, L$\alpha$
fluorescence requires a thorough treatment of the line transfer 
taking into account
partial redistribution.

\begin{table*}
\caption{Comparison between models and observations } \bigskip 
\begin{tabular}{ccccccc}
\hline
\noalign{\smallskip}
& model & model & model & mixed & composite & 1 Zw1 \\ & 1 & 
2 & 3 & model
& QSO	&	\\
\noalign{\smallskip}
\hline
\noalign{\smallskip}
L$\alpha$/H$\beta$ & 1.4 & 8 & 8 & 5 & 4.5	& 5 \\ 
FeIIUV/L$\alpha$ & 0.55 & 0.4& 0.64 & 0.47& 0.7	& 1 \\ 
FeIIopt/H$\beta$ & 3.6 & 1.8& 2.8 & 2.6 & $\geq$2.6 & $\geq$6.7 
\\
FeIIUV/Fe~IIopt	& 0.2 & 1.8& 1.8 & 0.8 & 1.3	& 0.7 \\ 
FeIItot/H$\beta$ & 5.4 & 5. & 8. & 5.1 & $\geq$6. & $\geq$11. \\ 
\noalign{\smallskip}
\hline
\end{tabular}
\medskip

{\it{Model 1: Collin et al. (1988), with low excitation, hydrostatic 
equilibrium as in an accretion disc, $\tau$(Bac)=4; 

Models 2 and 3: Wills et al. (1985), with high excitation, constant 
pressure and iron abundance solar and 3 times solar, respectively; 

Mixed model: 45$\%$ Model 1 + 55$\%$ Model 2. }} \bigskip 
\end{table*}

	Table 2 compares the observations with three photoionization 
models providing the strongest Fe~II lines. For all models, Fe~II 
emission is weak compared to that observed in 1 Zw1, the
prototype of the NLS1 class.

One should thus ask: WHAT WOULD BE THE NECESSARY 
INGREDIENTS OF A GOOD MODEL? Obviously one needs:
\begin{itemize}
\item a large column density of Fe$^+$
to get a large number of scatterings, in order to transform UV lines 
into optical lines ($\tau(2343) > 10^3$),
\item an efficient heating mechanism in the Fe$^+$ region to get 
collisional excitation up to 5eV levels ($T>$5000K) \end{itemize}
Let us see whether it would be possible to reach these conditions by 
playing with the parameters:
\begin{itemize}
\item influence of the ionizing-heating continuum: a strong IR and a 
strong hard X-ray continuum up to gamma rays would provide 
more heating to the HI$^*$ region, but it is not observed;
\item influence of the ionization parameter: it is strongly 
constrained, as it can be neither too high (it would lead to too 
intense L$\alpha$ emission, etc...), nor too low (it would not provide 
enough heating); 	 \item influence of the density: a high 
density would be favorable (to get a large Balmer opacity and to 
destroy Fe~IIUV), but it should not be too high (the Balmer and 
Paschen continua would then be too large); \item influence of the 
column density: it should be $\gg 10^{21}$ cm$^{-2}$ and it can be 
arbitrarily large, but this will not increase the amount of Fe~II 
emission above a given limit (since $T$ falls rapidly below the 
value required for Fe~II excitation); \item influence of the turbulent 
velocity: only a very high non-physical turbulence would allow an 
important fluorescence effect, and moreover it would decrease  
$\tau(2343)$ too much;
\item influence of the abundance of Fe: it is very weak, because of 
the thermostatic effect of Fe~II; an overabundance of a factor 10 
induces less than a factor 2 increase in the total flux; \item 
influence of the number of levels: it is not important for the total 
Fe~II emission which saturates (also the thermostatic effect) 
\end{itemize}
It is thus clear that {\it photoionization models cannot account for 
the Fe~II emission}, and the reason is that the region emitting these 
lines is the weakly ionized fraction of the clouds (the HI$^*$ region), 
where only a small amount of radiative flux can be absorbed per 
unit volume, owing to the small value of the photoionization cross 
sections in the soft X-ray range. For instance, in Verner et al. (1999), 
one sees that only 3$\%$ of the flux is emitted by this region. It also 
explains why {\it an overabundance of iron cannot help to increase 
the Fe~II emission in a photoionization model}, simply because it 
saturates at
the level of the radiative heating. One concludes that {\it the Fe~II 
region should be heated
by an additional non radiative
mechanism}. It is worth noting that the energy problem cannot
be solved by a strong reddening, as it would not reduce the 
discrepancy in
the ratio Fe~IIopt/H$\beta$ (cf. the controversy between Netzer, 
1985, Collin, 1986).

\subsection{Other models}

\begin{figure}
\centerline{\psfig{figure=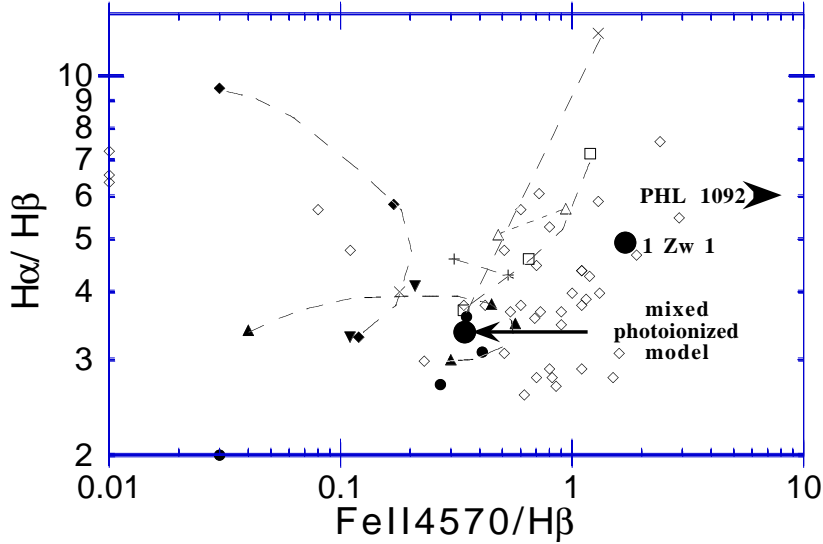,width=.8\textwidth}} 
\caption{Observed and computed line ratios: the diamonds are 
observed quantities, and the dotted lines represent collisional
models with various values of the temperature and of the density, 
and a variable column density: small filled circles: $T$=10000K, 
$n=10^{11}$cm$^{-3}$; filled diamonds: $T$=10000K, 
$n=10^{10}$cm$^{-3}$; filled triangles: $T$=8000K, 
$n=10^{12}$cm$^{-3}$; open squares: $T$=7000K, $n=10^{12}$cm$^{-
3}$; crosses: $T$=7000K, $n=10^{11}$cm$^{-3}$.}
\label{fig-Ha/Hb-4570/Hb}
\end{figure}

There are actually no models other than photoionization models, which 
require knowledge of the physics of the emission region. These are 
simply computations of the Fe~II spectrum assuming collisional 
ionization equilibrium at a given temperature (Joly, 1987). In these 
computations, the amount of energy available to heat the medium is 
arbitrary, so it allows us to get large Fe~II intensities. However, even in 
these computations, the parameters are constrained, since they 
could lead to anomalous line ratios. As an illustration we give in 
Fig. \ref{fig-Ha/Hb-4570/Hb} the observed and computed 
H$\alpha$/H$\beta$ ratio as a function of the Fe~II4570/H$\beta$ 
ratio (computations are from Joly, 1987). This figure
shows that a small temperature ($\sim$ 7000K) is required to 
account for the large Fe~II4570/H$\beta$ ratios, and still
the largest observed Fe~II4570/H$\beta$ ratios are not accounted 
for, because they would
correspond to too large H$\alpha$/H$\beta$ ratios. In these 
computations, the Fe abundance is normal. As the lines are optically 
thick, they are generally in the Voigt part of the curve of growth, so 
one expects their intensity to scale roughly with the square root of 
the abundance. In conclusion, a good fit would be obtained for a 
temperature of the order of 7000K, if one allows for an 
overabundance of a few units. 

\section{An empirical view}

	Other objects display intense Fe~II lines: 
cool luminous stars like Miras, with winds and chromospheres, 
which
are ``non-coronal" stars;
cataclysmic binaries, where it is suggested that Fe~II is formed in the
accretion disc, and novae in which Fe~II peaks soon after the 
outburst,
contrary to highly ionized species;
symbiotic stars like RR Tel, z Aur/W Cep, where Fe~II lines might be 
formed either in the accretion disc or in the cool component wind;
luminous hot stars like PCyg and B[e] supergiants with extreme 
stellar winds and accretion discs;
some type II supernovae.
	{\it These stars are characterized either by a strong 
variability, or by strong outflows, and
generally they have no X-ray spectrum and little or no ionizing 
continuum}. It is therefore tempting to extend these characteristics 
to the Fe~II emission regions of NLS1s. First, it would agree with the 
conclusion that the Fe~II spectrum is not produced directly by 
photoionization but more probably in shocks. Second, it leads to a 
consistent scenario which matches reasonably well the present 
picture of NLS1s, and which could explain the above mentioned 
correlations. 

One of the properties of NLS1s, often discussed in this meeting, is 
that they probably harbor relatively small black holes, accreting 
close to their Eddington rate. A large accretion rate means an 
increase of the fueling. It can be provided by
non-axisymmetric perturbations, for instance the presence of a 
bar, which trigger  an episode of star formation.
A strong accretion phase is therefore most probably linked with the 
existence of a nuclear starburst. This should be compared to high 
redshift luminous quasars, where heavy element overabundances 
are deduced in the BLR, and (for
BAL quasars) in the absorption line region, and are attributed to the 
presence of a starburst (Hamann \& Ferland, 1993, Collin \& Zahn, 
1999). 

The NLS1 characteristics might derive from this episode of star 
formation 
coupled with the high accretion rate.
Some enrichment in Fe is likely to occur. Strong outflows (due to 
stars and/or to a super Eddington
disc wind) could induce shocks and produce non radiative heating. 
The existence of outflows in NLS1s is also
strongly suggested
by the blue asymmetry of H$\beta$ (and of high ionization lines, as 
shown in this conference).
Due to the strong outflow, the NLR
might be replaced by a denser medium, thus explaining the 
weakness of the forbidden lines. The X-ray spectral properties 
(steepness and variability of the continuum) are generally 
attributed to NLS1s having their accretion rate in a
``high state phase", similar to that of galactic binaries. Concerning 
the ``narrowness" of the lines, several explanations have been 
proposed in this workshop, in particular linked with the spectral 
distribution. Collin and Hur\'e (in preparation) suggest that it could 
be simply due to a gravitational instability taking place in the 
accretion disk at a larger distance from the black hole in NLS1s than 
in normal Seyfert 1. Finally the anticorrelation of the Fe~II line 
intensities with the radio power could be due to the fact that only 
spiral galaxies are able to provide the nucleus 
with enough gas to reach a high accretion rate. 



\end{document}